\documentclass[a4paper]{jpconf}
\usepackage{graphicx}
\begin{document}
\title{Newton force from wave function collapse: speculation and test}

\author{Lajos Di\'osi}

\address{Wigner Research Center for Physics, Budapest 114, P.O.Box 49, H-1525 Hungary}

\ead{diosi.lajos@wigner.mta.hu}

\begin{abstract}
The Diosi-Penrose model of quantum-classical boundary postulates
gravity-related spontaneous wave function collapse of massive degrees
of freedom. The decoherence effects of the collapses are in principle
detectable if not masked by the overwhelming environmental decoherence. 
But the DP (or any other, like GRW, CSL) spontaneous collapses 
are not detectable themselves, they are merely the redundant formalism 
of spontaneous decoherence. To let DP collapses become testable physics, 
recently we extended the DP model and proposed that DP collapses are 
responsible for the emergence of the Newton gravitational force between 
massive objects. We identified the collapse rate, possibly of the order 
of 1/ms, with the rate of emergence of the Newton force. A simple heuristic 
emergence (delay) time was added to the Newton law of gravity. This 
non-relativistic delay is in peaceful coexistence with Einstein's 
relativistic theory of gravitation, at least no experimental evidence 
has so far surfaced against it. We derive new predictions of such a `lazy' 
Newton law that will enable decisive laboratory tests with available 
technologies. The simple equation of 'lazy' Newton law deserves theoretical 
and experimental studies in itself, independently of the underlying quantum 
foundational considerations. 
\end{abstract}

\section{Quantum mechanics of massive degrees of freedom}
Apart from the troublesome relativistic quantum gravity, we have no reason 
to doubt quantum mechanics' validity in the macro-world. 
If quantum mechanics is universally valid, massive d.o.f. of a bulk 
system must be quantized together with the light and atomic d.o.f.
For massive d.o.f., we should consider hydrodynamic modes 
in general. Unfortunately, the existence of large quantum uncertainties 
of massive d.o.f. in the so-called Schr\"odinger Cat states seems 
controversial. In a cautious modification of quantum mechanics 
\cite{Dio86,Dio87,Dio89,Pen94,Pen96,Pen98,Pen04}, parametrized by the
Newton constant $G$, spontaneous wave function collapses of massive d.o.f. 
will forbid the existence of Schr\"odinger Cats. A lasting problem
has been accompanying this (an other) spontaneous collapse models.
Spontaneous collapse is not testable, being the redundant formalism of 
spontaneous decoherence. What extension of the spontaneous collapse
theory makes it testable? --- we'll see below.
   
\subsection{Schr\"odinger Cats}
We don't discuss the quantum mechanics of hydrodynamical d.o.f. in general,
we consider the c.o.m. motion of a single spherical macroscopic object. In free
space, the wave function $\Psi(\mathrm{c.o.m.})$ satisfies the free Schr\"odinger 
equation. Accordingly, a well-localized initial $\Psi_0(\mathrm{c.o.m.})$ is 
unlimitedly expanding. Nothing prevents it to develop a macroscopic
wave packet, called the Schr\"odinger Cat (Fig.~1).
\begin{figure}[tbp]
\begin{center}
    \includegraphics[width=0.8\textwidth, angle=0]{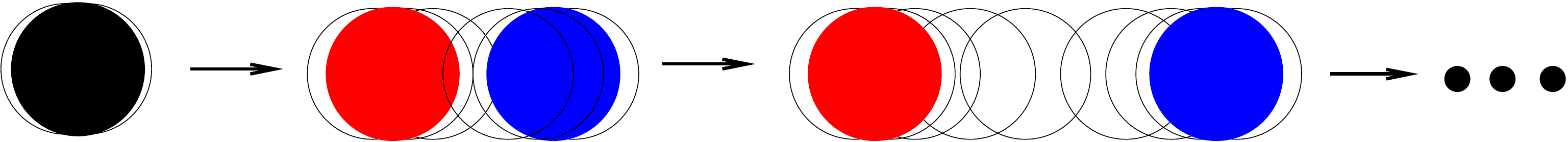}
\caption{Unlimited macroscopic quantum expansion of the c.o.m. of an isolated massive object.}
\end{center}
\end{figure}
The interactions with light d.o.f., e.g. with the surrounding thermal gas, 
will no way block the unlimited expansion of the c.o.m. quantum 
uncertainty. On the contrary, the composite wave function
$\Psi(\mathrm{c.o.m.},\mathrm{gas})$ will approximately follow the free 
Schr\"odinger equation in the c.o.m. \cite{Dio14x} 
while the c.o.m. is getting entangled with the gas (Fig.~2).
\begin{figure}[h]
\begin{center}
    \includegraphics[width=0.8\textwidth, angle=0]{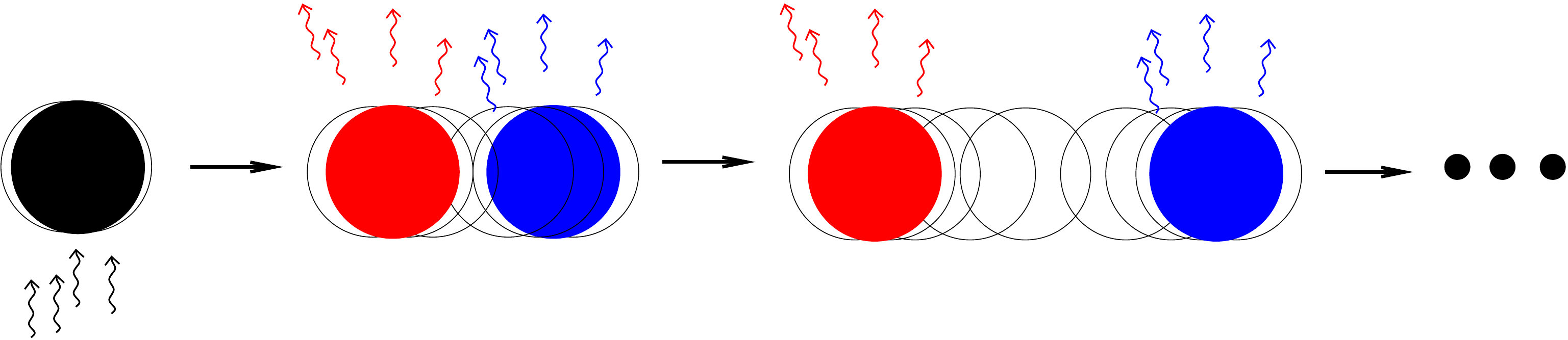}
\caption{Unlimited macroscopic quantum expansion of the c.o.m. of a massive object entangled with light environment.}
\end{center}
\end{figure}

Schr\"odinger Cats are problematic. Some consider the problems are merely 
metaphysical, not influencing the perfect consistency of quantum mechanics
as a physical theory. For others, including the author, the problem grows
physical because the von Neumann collapse of Schr\"odinger Cats provoke 
macroscopic violation of all basic conservation laws \cite{Dio14}.

Schr\"odinger Cats can be suppressed by spontaneous collapse models.

\subsection{G-related DP spontaneous collapse}
Many advocates and critics of different spontaneous collapse models
including GRW \cite{GRW86} and CSL \cite{CSL90} as well as DP, 
would overinterpret their subject constructing complicated notions. 
We rather propose a shortest definition in standard terms saving the 
reader's time and patience. 

In every spontaneous collapse model, the spontaneous collapses
are caused by standard von Neumann measurements with three specific features.
\begin{itemize}
\item{The devices are weakly measuring certain universal observables.} 
\item{The devices are present everywhere and every time.} 
\item{The devices are hidden, they decohere all accessible systems yet the
measurement outcomes remain inaccessible.}
\end{itemize}
Specifically, in DP theory the hidden devices are universally measuring the 
local mass density operator of our physical system, the strength parameter
of the measurement is scaled by the Newton constant $G$. Accordingly,
the local mass density is universally enduring random collapses (disentanglement)
by the hidden devices that forbid macroscopic quantum uncertainties of massive
d.o.f. and, in particular, prevent Schr\"odinger Cats of being created. 
\begin{figure}[h]
\begin{center}
    \includegraphics[width=0.80\textwidth, angle=0]{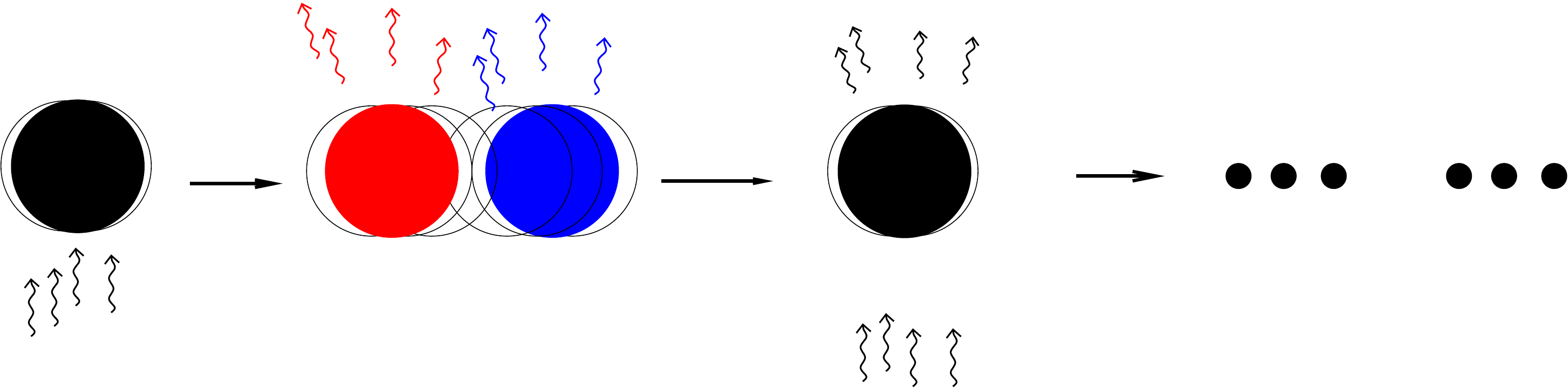}
\caption{Spontaneous collapse (disentanglement) of the massive d.o.f., leading
to random expansion-collapse cycles of characteristic time $\tau_G$.}
\end{center}
\end{figure}

Under natural circumstances, unitary evolution by the Schr\"odinger
equation and the random DP spontaneous collapses compete with each other.
The unitary evolution makes the c.o.m. uncertainty grow whereas the
collapses make it shrink. The two can reach balance at a certain
equilibrium collapse rate $1/\tau_G$ (where $\tau_G$ stands for the
equilibrium collapse time scale). The following estimation can be
derived:
\begin{equation}
\frac{1}{\tau_G}\sim\sqrt{G\rho^\mathrm{nucl}}\sim\frac{1}{\mathrm{millisecond}}
\end{equation}
where $\rho^\mathrm{nucl}$ is the nuclear mass density. Details of the
estimation, and the particular reasons of the appearance of the 
nuclear mass density in a literally non-relativistic model, are given
in Refs. \cite{Dio13a,Dio14}. We can conclude that in DP theory the
massive (hydrodynamic) d.o.f. are, under natural conditions, enduring 
spontaneous collapses at time scales $\tau_G\sim1$ms (Fig.~3).

\subsection{Decoherence vs collapse}
According to spontaneous collapse theories, the spontaneous
decoherence effects are detectable in principle. Experiments are under way 
\cite{Mar_Bou03,Van_Asp11,Rom11,LiKheRai11,Pep_Bou12}
to suppress environmental decoherence that always masks spontaneous
decoherence unless we get our massive d.o.f. extreme isolated.
Most theoreticians, therefore some experimentalists as well, 
don't emphasize (or even fail to recognize) that the spontaneous
collapse theories are redundant realizations of the corresponding spontaneous decoherence
theories. Spontaneous collapse (disentanglement) is never detectable, only the resulting
spontaneous decoherence is, see \cite{Dio89}. This becomes obvious from our previous
definition of spontaneous collapse theories. Since the hypothetic
universal measuring devices are hidden, the measurement outcomes are hidden,
one can never detect the collapse of the wave function. 
What one could detect is the damping of certain well-defined interference terms, 
i.e.: the phenomenon of spontaneous decoherence.
\begin{figure}[tbp]
\begin{center}
    \includegraphics[width=0.80\textwidth, angle=0]{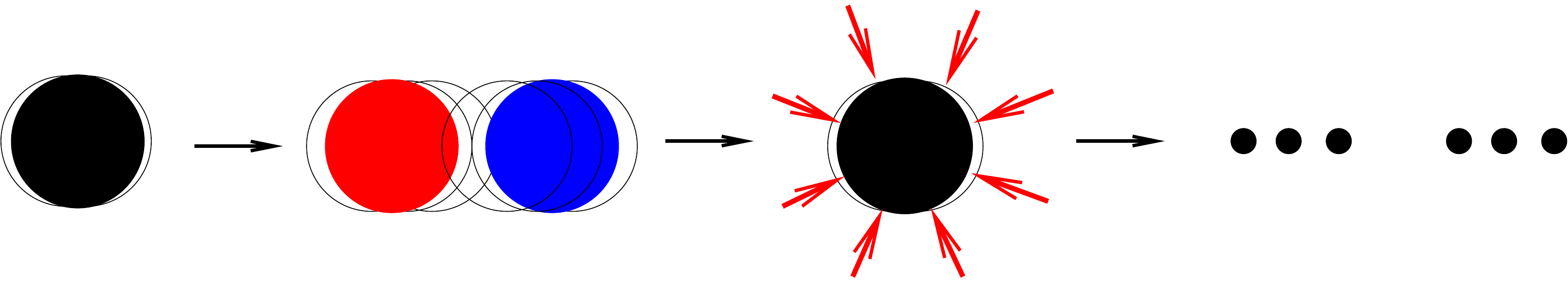}
\caption{Random spontaneous collapse of massive d.o.f. triggers the
Newton gravitational field at a characteristic emergence time $\tau_G$.}
\end{center}
\end{figure}

To let collapse become testable physics, one has to associate it with something physical! 
Recently we have extended the DP theory, supposed that the Newton force is 
generated by DP collapses. How should the Newton field emerge from DP collapses? 
Diverse and highly heuristic arguments have appeared in earlier works \cite{Dio07,Dio09}. 
But a detailed model is missing. In fact, we talk about an assumption beyond the DP theory,
in minimal formulation.

Suppose the G-related DP collapses induce Newton gravity. Then, independently of 
the detailed mechanism, the emergence rate/time of Newton gravity is related 
(proportional) to the wave function collapse rate/time of the sources. Actually,
the emergence time scale of Newton gravity must be of the order of the collapse
time scale $\tau_G\sim1$ms (Fig.~4).
We outline how this hypothesis would lead to a particular modification of the standard Newton theory,
offering robust effects compared to an environmentally masked spontaneous decoherence.

\section{Lazy Newton force with a delay}
Unlike quantum foundational considerations in the previous Section, 
our resulting task is purely classical, i.e., non-quantum. 
We have to modify the purely non-relativistic Newton's law to account
for a purely non-relativistic delay time $\tau_G$ which we think to be of the order of a millisecond. 
A minimum naive extension of Newton's classical equation would be 
\begin{equation}\label{LazNew}
\phi(r,t)=\int_0^\infty\frac{-GM}{\vert r-x_{t-\tau}\vert}\mathrm{e}^{-\tau/\tau_G}\frac{d\tau}{\tau_G}
\end{equation}
where $\phi(r,t)$ is the Newton potential at location $r$ and time $t$, of a source 
of mass $M$ moving along the trajectory $x_t$. The field at time $t$ depends on the 
position of the source then and before, including a period of the order of $\tau_G$.
For zero delay $\tau_G=0$, our Eq.~(\ref{LazNew}) restores Newton's. 

Unfortunately, for $\tau_G>0$ the predicted modification would violate
the equivalence of inertial frames and the equivalence of accelerated frames
with gravity. 
Consider a source at rest at $x_t\equiv0$, yielding $\phi(r,t)\equiv-GM/r$. Let's
boost our inertial frame by a constant velocity $v$ so that $x_t=-vt$ in the new
inertial frame where (\ref{LazNew}) yields
\begin{equation}
\phi(r,0)=\int_0^\infty\frac{-GM}{\vert r-v\tau\vert}\;\mathrm{e}^{-\tau/\tau_G}\frac{d\tau}{\tau_G}\;.
\end{equation}
The gravitational field $\phi(r,0)$ has changed from one inertial frame to
the other. This shows that our naive equation (\ref{LazNew}) is not consistent in itself. 
Without illustrating how the naive equation (\ref{LazNew}) violates also the equivalence
of accelerated frames and gravity, we show the resolution below.

In Ref.~\cite{Dio13b}, we proposed that the naive equation (\ref{LazNew}) is valid and should be
applied in the co-moving-free-falling reference frame. In other words, before
we calculate the Newton potential of the moving source at time $t$, we install an instantaneous
reference frame where $\dot x_t=0$ and $M\ddot x_t$ is equal to the 
non-gravitational force if there is any. Having calculated $\phi(r,t)$ in
the co-moving-free-falling frame, we can transform the result back to any other frame.

Hence we guarantee that the proposed modification of the Newton law is the
same in all inertial frames. One can also show that the equivalence between
gravity and an accelerated reference frame is respected. We should emphasize
a major consequence of our chosen frame for Eq.~(\ref{LazNew}): 
Newton law is completely restored in absence of non-gravitational forces.  
Accordingly, the effect of the delay time $\tau_G$ can only be expected 
with a massive source which is under the influence of a non-gravitational
force. The point is not that the source is moving or isn't, the point is
the presence of a non-gravitational force. We show three different 
experimental situations.

\subsection{Testable predictions of gravity's laziness I.}
Interestingly, there is a universal effect in Earth gravitational acceleration
$g=9.81\mathrm{cm/s^2}$. 
Let start from free falling objects first and apply Eq.~(\ref{LazNew}) in the 
co-moving-free-falling frame as we have to. The result is that the free-falling
object creates the standard instantaneous Newton potential. Now we consider
an object at rest in the laboratory frame. Again, we use the co-moving-free-falling
frame where the object is accelerated by the non-gravitational force through the
support or the thread. We can calculate the effect of the delay in (\ref{LazNew})
in the lowest order of the delay time $\tau_G$, leading to a surprising result. 
\begin{figure}[h]
\begin{center}
    \includegraphics[width=0.75\textwidth, angle=0]{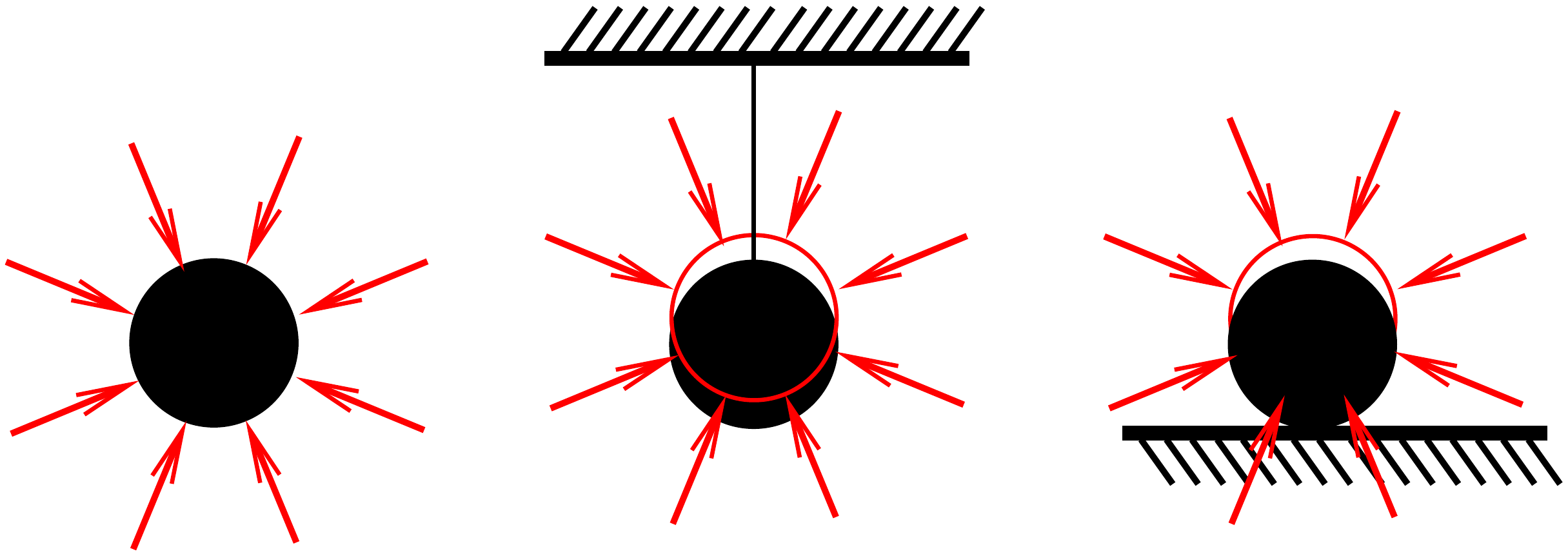}
\end{center}
\caption{Free-falling object (left) in Earth gravity creates standard Newton field. 
Static sources, suspended (middle) or supported (right) create standard Newton field 
shifted slightly upward by a universal height $\delta_G=g\tau_G^2$.}
\end{figure}
All static (supported or suspended) objects create Newton forces as if these
objects were higher than their geometric position (Fig.~5), by a universal height
\begin{equation}\label{dGh}
\delta_G=g\tau_G^2\;.
\end{equation}
For $\tau_G\sim1$ms this universal height is of the order of $10^{-3}$cm (or
nearly $10^{-2}$cm), a surprisingly big value. 
In a recent Cavendish experiment the positioning uncertainty of the source 
and test masses is cca. $0.5\times\mu$m \cite{Quietal13}. 
This would enable detecting a $10^{-3}$cm vertical (there lateral) shift of the 
source's field.

\subsection{Testable predictions of gravity's laziness II.}
Alternative to static laboratory objects in Earth gravitational field, 
also ones accelerated by non-gravitational forces can test  
the delay $\tau_G$. As a typical example, consider a source revolving at small angular 
frequency $\Omega\ll 1/\tau_G$ on a rope of length $R$. 
For simplicity, we disregard the Earth gravity and
calculate the effect of mechanical acceleration $R\Omega^2$ only. Applying Eq.~(\ref{LazNew}) 
in the instantaneous co-moving frame we find that the source creates Newton
field as if it were closer than $R$ to the center by $\delta_G$. 
A lowest order expansion yields a similar result to (\ref{dGh}), with $R\Omega^2$ 
in place of $g$:
\begin{equation}\label{dGR}
\delta_G=R\Omega^2\tau_G^2\;,
\end{equation}
valid if $\Omega\ll 1/\tau_G$. Suppose we are interested in the Newton field in the center
of the source's orbit. Since the source generates the field from a shorter distance
$R-\delta_G$, we get stronger Newton field by the factor $1+\Omega^2\tau_G^2$ \cite{Err},
a small effect for slow revolution and small acceleration (Fig.~6).  
Experiments with revolving sources, like \cite{GunMer00}, might make decisive tests or
put an upper limit on $\tau_G$ at least.
\begin{figure}[h]
\begin{center}
    \includegraphics[width=0.75\textwidth, angle=0]{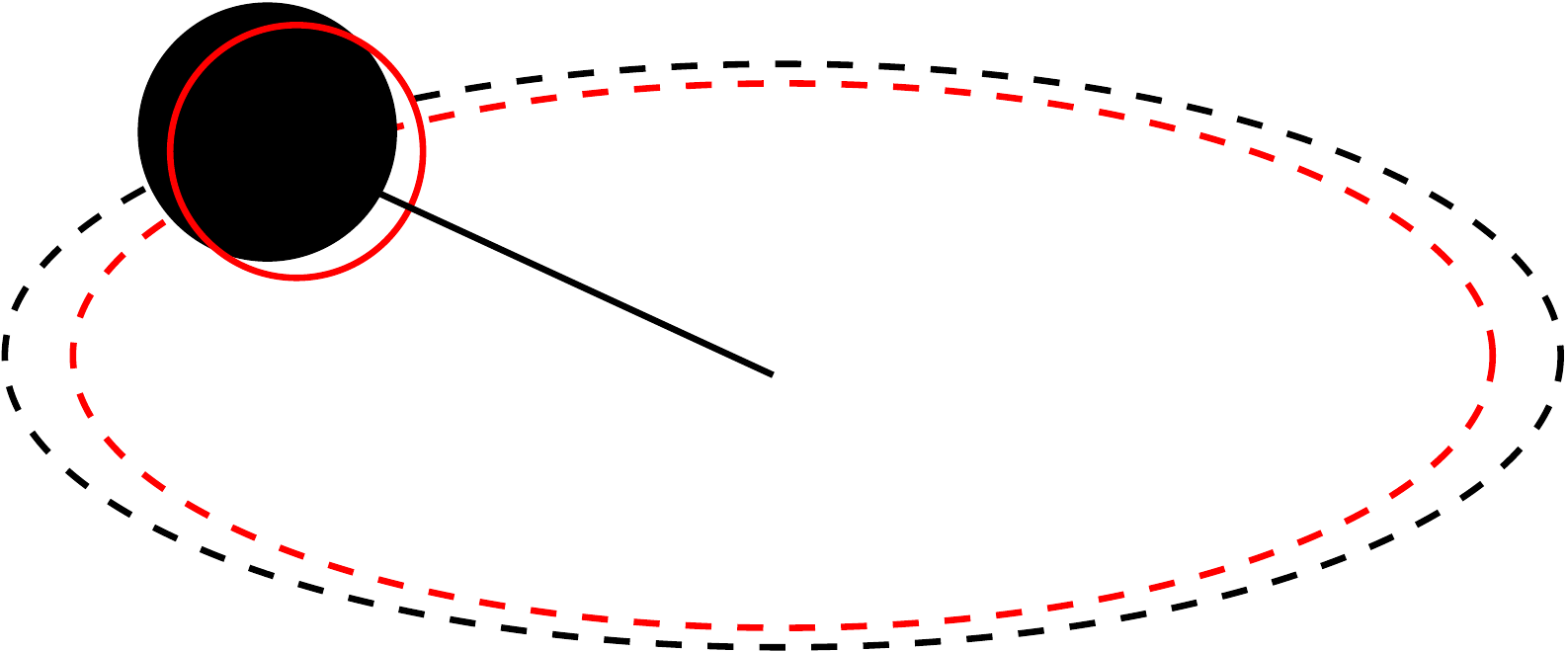}
\caption{Revolving sources at angular frequency $\Omega\ll 1/\tau_G$ on a circle of radius $R$ 
create standard Newton field as if the radius were smaller by a factor $1-\Omega^2\tau_G^2$.}
\end{center}
\end{figure}

\subsection{Testable predictions of gravity's laziness III.}
Direct test of the delay $\tau_G$ is possible if we impose extreme non-gravitational acceleration
on our source. Let us prepare a massive laboratory source at static location $x_t\equiv0$ for $t<0$. 
It develops the standard static Newton field $-GM/r$ for $t<0$, apart from the tiny universal shift 
$\delta_G$ (\ref{dGh}) discussed previously. Here, again, we disregard this small effect, rather we look 
for a robust explicit delay of the Newton field. To this end, at $t=0$ we suddenly displace the object to a 
new static position $x_t\equiv a$ for $t>0$. Due to the static sources, and because we ignore the
effect of the Earth gravity, we can apply Eq.~(\ref{LazNew}) in the laboratory system, yielding
\begin{equation}
\phi(r,t)=\mathrm{e}^{-t/\tau_G}\frac{-GM}{r}
          +\left(1-\mathrm{e}^{-t/\tau_G}\right)\frac{-GM}{\vert r-a\vert} 
\end{equation}
for $t>0$. From the old to the new Newton fields around respectively the old
and new locations $0$ and $a$, the change is gradually taking place
with a characteristic delay time $\tau_G$. This is exactly what we expect of the
modified Newton law (\ref{LazNew}): the lazy Newton field follows the sudden displacement of
the source with the delay $\tau_G$ (Fig.~7).
\begin{figure}[h]
\begin{center}
    \includegraphics[width=0.75\textwidth, angle=0]{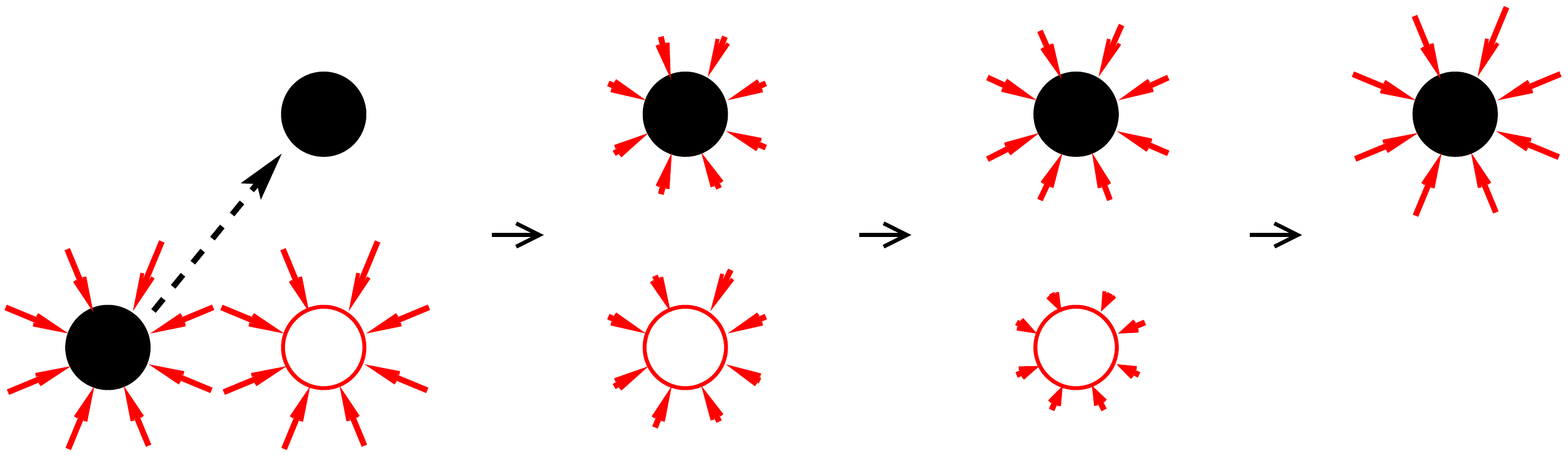}
\end{center}
\caption{Sudden displacement of the source followed by the delayed redistribution of the Newton
field from the old to the new configuration.}
\end{figure}
In a real well-controlled test, only small macroscopic objects can be displaced or removed 
faster than a millisecond. Short-distance gravimetry at fine time-resolution is requesting
but not impossible.

\section{Summary, outlook}
In the first part, we outlined the DP theory of spontaneous collapse of massive d.o.f.,
yielding an equilibrium time scale $\tau_G\sim1$ms of the collapse (disentanglement). 
We emphasized that in spontaneous collapse theories, like DP and all the others, the spontaneous decoherence is
the testable effect, the spontaneous collapse (disentanglement) is not. We invoked the recent 
extension of DP theory \cite{Dio13b} that makes collapse testable by associating it with the emergence
of the Newton potential. 
The second part discussed the heuristic modification of Newton law to incorporate the `lazy'
emergence of the Newton potential just with the delay time of the order of $\tau_G$. We showed
three different laboratory tests of the delay. The first test is based on the indirect effect of the delay, 
that static Newton fields of static sources in Earth gravitational field will get shifted upward by 
a universal height. A second test may exploit the slight modification of the standard Newton field 
of a source in periodic orbital motion. A direct test of the delay is possible with the sudden 
(i.e.: faster than $\tau_G$) displacement of a small macroscopic source where the old/new 
configurations should die/emerge exponentially with the characteristic time $\tau_G$. 
The realization of these tests seems possible with recent technologies.

We have emphasized that the concept of a lazy emergence of the Newton field is extreme speculative,
the model is heuristic and minimal. As a non-relativistic theory, it is in peaceful coexistence 
with the general relativistic theory of gravitation. On one hand experimental evidences have not yet 
surfaced against it, and on the other hand it has various detectable predictions.

The modified Newton law---given by Eq.~(\ref{LazNew}) in co-moving-free-falling frame---is a naive 
classical non-relativistic model, it can be considered both by theorists and experimentalists 
independently of the quantum mechanical DP theory, so that the value of the delay parameter 
$\tau_G$ could be tested in ranges different from the quantum theoretically predicted milliseconds. 

\ack
This work was supported by the Hungarian Scientific Research Fund under Grant No. 75129 and
the EU COST Action MP1006 `Fundamental Problems in Quantum Physics'. The author is indebted 
to the organizers of EmQM13 for the generous invitation.


\begin{thebibliography}{9}
\bibitem{Err} Erratum: factor $1/2$ was redundant in the Taylor-expansion (4) in \cite{Dio13b}, also led to a redundant $1/2$ within the enhancement factor there.
\end{thebibliography}
\end{document}